# Topological nanophononic states by band inversion

*Martin Esmann,[1,‡] Fabrice Roland Lamberti,[1,2,‡] Pascale Senellart,[1] Ivan Favero,[2] Olivier Krebs,[1] Loïc Lanco,[1,3] Carmen Gomez Carbonell,[1] Aristide Lemaître,[1] and Norberto Daniel Lanzillotti-Kimura[1,*]*

[1] Centre de Nanosciences et de Nanotechnologies, CNRS, Université Paris-Sud, Université Paris-Saclay, C2N Marcoussis, 91460 Marcoussis, France

[2] Matériaux et Phénomènes Quantiques, Université Paris-Diderot, CNRS UMR 7162, Sorbonne Paris Cité, 10 Rue Alice Domon et Léonie Duquet, 75013 Paris, France

[3] Université Paris Diderot Paris 7, 75205 Paris CEDEX 13, France

‡These authors contributed equally.

**Abstract**

Nanophononics is essential for the engineering of thermal transport in nanostructured electronic devices, it greatly facilitates the manipulation of mechanical resonators in the quantum regime, and could unveil a new route in quantum communications using phonons as carriers of information. Acoustic phonons also constitute a versatile platform for the study of fundamental wave dynamics, including Bloch oscillations, Wannier Stark ladders and other localization phenomena.

Many of the phenomena studied in nanophononics were indeed inspired by their counterparts in optics and electronics. In these fields, the consideration of topological invariants to control wave dynamics has already had a great impact for the generation of robust confined states. Interestingly, the use of topological phases to engineer nanophononic devices remains an unexplored and promising field. Conversely, the use of acoustic phonons could constitute a rich platform to study topological states.

Here, we introduce the concept of topological invariants to nanophononics and experimentally implement a nanophononic system supporting a robust topological interface state at 350 GHz. The state is constructed through band inversion, i.e. by concatenating two semiconductor superlattices with inverted spatial mode symmetries. The existence of this state is purely determined by the Zak phases of the constituent superlattices, i.e. that one-dimensional Berry phase. We experimentally evidenced the mode through Raman spectroscopy. The reported robust topological interface states could become part of nanophononic devices requiring resonant structures such as sensors or phonon lasers.

**Introduction**

In macroscopic acoustics exciting effects such as acoustic cloaking [1,2] superlensing [3] traps for electrons [4] and rainbow trapping [5] have recently been reported. Nanophononics, relying on the same wave mechanics, addresses the engineering and manipulation of high frequency phonons at the nanoscale [6–9]. Phonon engineering in the GHz/THz range has major implications in other domains: in optomechanics for the manipulation of mechanical resonators in their quantum ground state [10,11], in electronics for determining the thermal transport properties of nanostructured devices[8,12,13] and even in solid-state quantum communications, where acoustic phonons could serve as carriers of quantum information [14–16]. In the high frequency regime, the resulting low thermal occupation number allows to readily prepare mechanical systems in their quantum regime [10,11].

Fundamental building blocks in nanophononics are finite size nanoscale superlattices [17,18] presenting a periodic modulation of the elastic properties. Such devices exhibit high reflectivity bands for acoustic phonons in the GHz-THz range and are usually employed as distributed Bragg reflectors (DBRs) [19,20]. DBRs are at the heart of key advances in nanophononics such as acoustic nanocavities formed by enclosing a resonant acoustic spacer in between two DBRs [9,19–23]. The high reflectivity bands of a phononic DBR originate directly from the associated acoustic minigaps of the corresponding infinite periodic superlattice.

The low speed of sound and the long mean free path of acoustic phonons make the full phononic wave function information accessible to optical probes. Together with state-of-the-art nanofabrication technologies, engineered acoustic phonons constitute a versatile platform for the investigation of complex wave dynamics and localization [21,24,25]. For instance, based on one-dimensional nanophononic structures impressive advances have been reported on the feeding of a laser mode by shaking quantum dots [26], on the development of efficient optomechanical platforms [11], and on coherent THz sound amplification [23]. However, up to now, the control of acoustic phonons propagation in periodic media solely relies on $\lambda/4$ interference stacks, creating frequency intervals where elastic waves cannot propagate, i.e. phononic bandgaps. The full information contained in the acoustic band diagram, in particular the underlying spatial mode symmetries, have not yet been exploited to design acoustic devices beyond the standard Fabry-Perot resonator.

Topological invariants have been widely used to describe the quantum Hall effect [27–30] and electrically conducting polymers [31,32] and for the conception of unidirectional optical waveguides [33,34]. In periodic media, topological invariants allow for an efficient description of the information beyond the mere bandgap existence. For one-dimensional systems, the Zak phase [35], i.e. the one-dimensional Berry phase [36], is usually invoked as a topological number [31,37–40]. For instance, the Zak phases corresponding to two concatenated systems determine the existence of an interface mode confined between them. Such a mode is robust against perturbations in the systems that do not affect the values of their Zak phases. Recently, first reports have merged these concepts from topology with acoustics in the kHz-MHz range [37,41–45].

In this work, we introduce the concept of topological invariants to nanomechanics in the hundreds of GHz range and experimentally implement a nanophononic system in which an interface state at 350GHz is constructed. This topological state is designed through band inversion [38], i.e. by concatenating two superlattices with inverted spatial mode symmetries at the band edges around a common minigap, in the absence of a resonant spacer. The existence of this state is purely determined by the Zak phases of the constituent superlattices. It is then experimentally evidenced through high resolution Raman spectroscopy. As the development of phonon lasers [23] and optomechanical sensing applications rely on resonant structures, robust topological interface states could become a powerful ingredient in the development of nanophononic devices.

**Results and Discussion**

The direct link between an acoustic DBR and the topological properties of its corresponding band structure constitutes the base of our study [46].Let us establish the connection between the topological properties of periodic media and traditional concepts in phononics and photonics. Since a DBR is a periodic medium for phonons, it has an associated band structure with frequency bands of propagating Bloch modes and band gaps, in which only evanescent phonons are solutions to the acoustic wave equation [46]. A state confined in between two concatenated DBRs can only exist for frequencies that fall into a band gap for both DBRs. To get a localized state in a cavity made by two DBRs and a spacer, i.e. a Fabry-Perot resonator [47], the reflection phases $\phi_{left}$ and $\phi_{right}$ of the individual reflectors and the phase picked up by propagation through the spacer have to add up to an integer multiple of $2\pi$, i.e. a stationary wave is formed according to

$$\phi_{left} + 2\phi_{spacer} + \phi_{right} = 2m\pi, \quad m \in \mathbb{Z} \qquad (1)$$

The DBR reflection phases can be positive or negative depending on the structure of the considered DBRs. In a more general picture, Eq. (1) can also be fulfilled in the extreme case of the complete absence of a spacer, when directly concatenating two different DBRs in order to generate an interface state, that is

$$\phi_{left} + \phi_{right} = 2m\pi, \quad m \in \mathbb{Z} \qquad (2)$$

It is noteworthy that this phase condition is general and therefore also applies to other systems, for example localized surface plasmons [48], electromagnetic waves pinned at the interface between two optical materials or an electronic wave localized at the interface between two semiconductors [49].

A yet unexplored way to fulfill Eq. (2) in nanophononics is by making use of topological properties related to infinite superlattices through the concept of band inversion. In two-dimensional materials the concept of inverted band structures usually refers to systems where the conduction and valence band symmetries are inverted [29]. In the context of this work, we denote that two one-dimensional systems present inverted bands when: 1) they have a common band gap; 2) the modes at the band edges present opposite spatial symmetries. These two systems belong to two different topological phases as discussed below. The intimate relation between topological phases and Eq. (2) arises since two DBRs with inverted bands present opposite signs in the reflection phases across the common minigap [37,38].

One of the simplest realizations of the band inversion principle is depicted in Figure 1. We consider a DBR which is constituted by alternating layers of GaAs and AlAs with acoustic impedances $Z_{GaAs} = \rho_{GaAs} v_{GaAs}$ and $Z_{AlAs} = \rho_{AlAs} v_{AlAs}$ ($\rho$ mass density and $v$ speed of sound). At a design frequency $f_0 = 175 GHz$ the total acoustic path length of the unit cell is set to half a phonon wavelength $\lambda/2$, i.e. the thicknesses $d$ of the two layers obey $\frac{d_{GaAs}}{v_{GaAs}} + \frac{d_{AlAs}}{v_{AlAs}} = \frac{1}{2f_0}$ and a phase of $\pi$ is accumulated by a phonon at frequency $f_0$ traversing both layers of the cell. As a consequence, all band gaps of the DBR are centered at integer multiples of $f_0$.

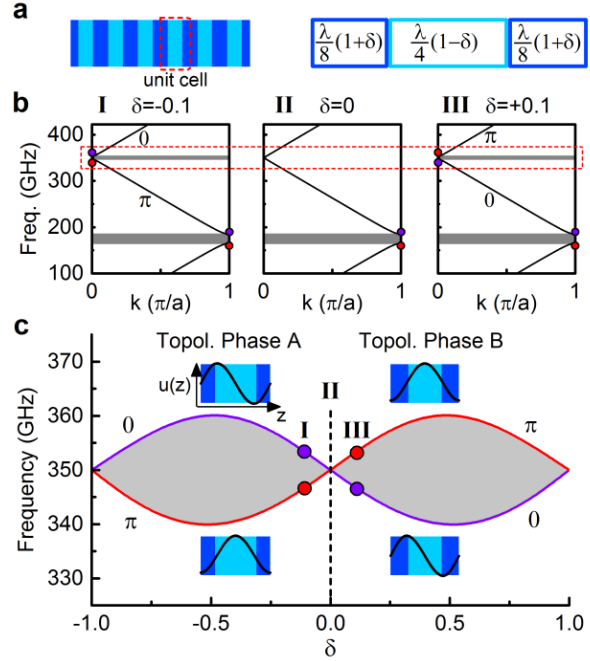

**Figure 1 Band inversion and topological phases of a nanophononic DBR, a,** Schematic of a nanophononic DBR and its unit cell by parametrized by δ which describes the relative thickness of the materials (see text). Dark (light) shades correspond to GaAs (AlAs) layers. **b,** Acoustic band structures of a nanophononic DBR for three different values of δ. The first and second minigap can be identified around 175GHz and 350GHz, respectively. For $\delta = 0$ (center) the second minigap is closed, while for $\delta = \pm 0.1$ (left and right) it is open. The symmetries of the modes at the Brillouin zone center (indicated with colored circles) are inverted in energy. Accordingly, the Zak phases of the two bands bounding the second minigap exchange [21]. **c,** Band inversion of the second acoustic minigap around 350GHz. Shown are the frequencies of the band edges (violet and red) bounding the minigap (grey) as a function of the parameter δ. A sign change in δ marks the transition between the topological phases A and B of a DBR. While for δ<0 the Bloch mode at the lower (upper) band edge has a symmetric (anti-symmetric) displacement pattern with respect to the centers of the material layers; these symmetries exchange for δ>0. The edge modes are illustrated in the insets of panel c.

To describe how the overall acoustic path length is distributed between the two materials, we define a parameter $-1 < \delta < 1$ as sketched in Figure 1a. Keeping $f_0$ constant, the thicknesses of the layers are $d_{GaAs} = \frac{v_{GaAs}}{4f_0}(1+\delta)$ and $d_{AlAs} = \frac{v_{AlAs}}{4f_0}(1-\delta)$. The particular case of a DBR made of $\lambda/4$ layers is therefore described by $\delta = 0$.

In Figure 1b we show three acoustic band structures corresponding to cases of different values for δ. First, for δ=-0.1 the second minigap is open, presenting a symmetric (anti-symmetric) Bloch mode at the lower (upper) band edge (see insets on the left of Figure 1c). Second, for δ =0 the second minigap is closed and thus no symmetries can be assigned to the degenerate edge modes. Third, for δ =+0.1 we observe the same bandgaps as in the first case, but the spatial symmetries of the band edge modes are inverted (see insets on the right of Figure 1c). We denote (anti-)symmetric modes with a violet (red) dot.

We can follow the evolution of the width of the acoustic minigap when varying the value of δ continuously (shown in Figure 1c). The violet and red lines indicate the frequencies of the two band edges enclosing the considered gap. In grey, the span of the minigap is indicated. Exactly at δ =0 the symmetries of the edge modes undergo an inversion, marking a topological transition. A topological transition is usually characterized by topological invariants such as the Zak phase (i.e., the Berry phase for Bloch bands in one dimension [35]). The Zak phase of the acoustic bands can be computed by an integration across the Brillouin zone as follows [37]:

$$\theta_{Zak}^n = \int_{-\pi/a}^{\pi/a} \left[ i \int_{unit\ cell} \frac{1}{2\rho(z)v(z)^2} dz u_{n,k}^*(z) \partial_k u_{n,k}(z) \right] dk \qquad (3)$$

Here, $u_{n,k}(z)$ is the mechanical displacement of the Bloch mode's cell-periodic part in the n-th band as a function of position $z$ along the superlattice.

As it can be observed in Figure 1, the Zak phases corresponding to the phononic bands bounding the second minigap from the top and below appear inverted in energy when crossing the topological transition point at δ=0. It has been demonstrated [38] that the Zak phases corresponding to the bands below a certain minigap are directly linked to the sign of reflection phase $\phi$ in that minigap, hence establishing the link with Eq. (2). For the second minigap the sign of the reflection phase is determined by the Zak phases of the zeroth and the first bands [38]

$$\text{sgn}[\phi] = \exp\left[i(\theta_{Zak}^0 + \theta_{Zak}^1)\right] \qquad (4)$$

Note that to establish the link between the band structure and the reflectivity properties of a DBR were established by assuming a semi-infinite DBR terminated at the center of a layer. Terminating the DBR at the center of a layer implies that the first unit cell is centro-symmetric as sketched in Figure 1a. In this case, the connection between mode symmetries and reflection phases can be heuristically understood, since the anti-node (node) at the DBR surface results in a reflection phase $\phi$ evolving from 0 to $\pi$ across the bandgap for $\delta < 0$ and from $-\pi$ to 0 for $\delta > 0$, respectively. That is, the sign of $\delta$ directly determines the sign of $\phi$ for frequencies inside the bandgap [38,46].

Consequently, by concatenating two DBRs with inverted bands as shown in Figure 2a, the first one with $\delta = -0.1$ and the second one with $\delta = +0.1$, the resonance condition Eq. (1) is automatically fulfilled in the second minigap. Notice that the two DBRs present exchanged spatial mode symmetries at the band edges. Likewise, in the left DBR the Zak phases of the bands bounding the second minigap are inverted with respect to the ones in the right DBR. Since the DBRs terminate at the center of a GaAs layer the resonance condition is fulfilled at the *center* of the minigap (See Appendix B). Figure 2b shows the calculated acoustic reflectivity for a structure composed of 20 unit cells in each DBR. A clear stop band and a dip in the reflectivity can be observed. The minigap shown in Figure 2a corresponds to the stop band in Figure 2b. In the limit of an infinite number of pairs in the DBRs the stop band coincides with the minigap. The dip at 350 GHz corresponds to the phononic mode confined at the interface between the two DBRs. We have calculated the corresponding displacement profile, i.e. the modulus of the mechanical displacement $|u(z)|$, and show it in Figure 2c (black) superimposed by the DBR layer schematics. Light and dark colors represent AlAs and GaAs layers, respectively. For clarity, green and blue indicate the two different topological phases. The envelope of the field has a maximum at the interface between the two DBRs with different topological phases. This mode disappears if the bands of the two constituent DBRs are not inverted (not shown here).

In nanophononic and nanophotonic applications, DBRs are usually constituted by an integer number of bilayers. That is, the array is terminated at the interface between two different materials. For instance, n periods of GaAs/AlAs bilayers, in contrast to what was shown in Figure 2, where an integer number of centro-symmetric unit cells is considered. As we will show, the bilayer approach also leads to topological interface states.

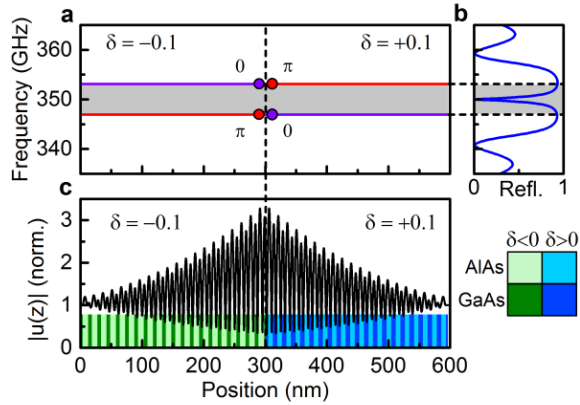

**Figure 2 Topological interface state at 350GHz;** a, Local phononic band diagram of two concatenated DBRs presenting inverted bands ($\delta = -0.1$ for the left DBR and $\delta = +0.1$ for the right DBR). Notice that the Zak phases and the mode symmetries are inverted at the interface. **b**, Phonon reflectivity corresponding to the structure indicated in a. Each DBR contains 20 centro-symmetric unit cells. The mode at 350GHz corresponds to the topologically confined state, which appears at the center of the acoustic minigap. **c**, Spatial displacement pattern $|u(z)|$ of the topological interface state at 350GHz (black) together with the DBR structure. The mode envelope shows a maximum at the interface between the two DBRs and decays evanescently into both directions away from the interface. Green and blue color schemes denote spatial regions with different topological phases.

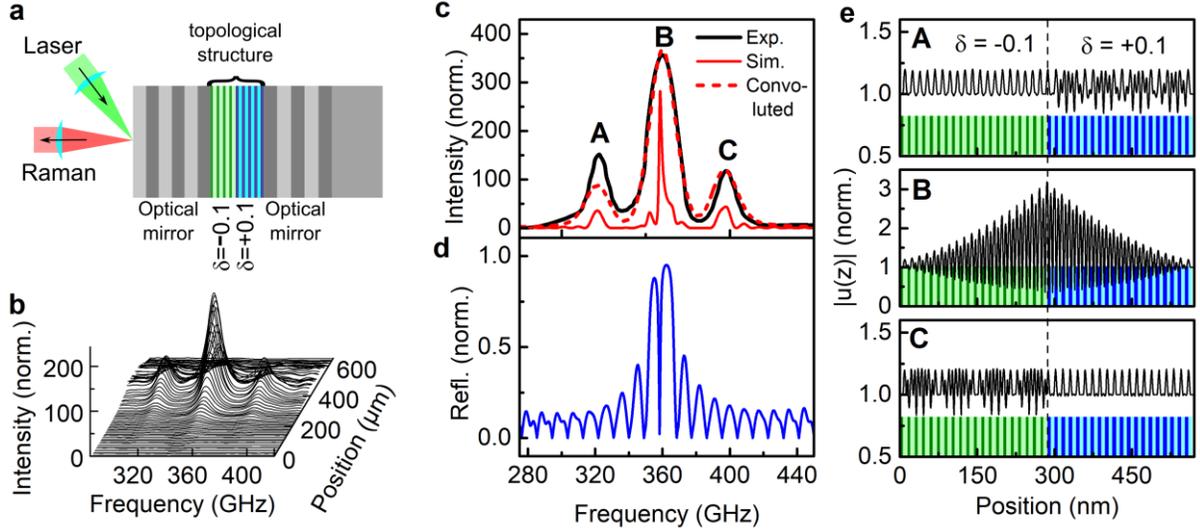

**Figure 3 High resolution Raman spectroscopy of a topological nanophononic interface state; a**, Sample structure with the topological acoustic structure acting as the 2λ wide spacer of an optical superstructure. **b**, Raman spectra as a function of laser incidence position on the sample showing the double optical resonance (DOR) condition. **c**, Experimental Raman spectrum (solid black) together with photoelastic model calculation (solid red). The model calculation is convoluted with a Gaussian (2σ = 13GHz) to account for the finite experimental resolution (dashed red). **d,** Simulated acoustic reflectivity of the sample. The topological interface mode at 360 GHz corresponds to peak B in the Raman spectrum. **e**, Layer schematics of the topological structure together with mechanical mode profiles corresponding to the peaks marked A through C in panel c. Peak B originates from the topological interface mode showing a maximum at the interface between the two superlattices, peaks A and C are extended modes in the structure.

To prove the existence of topological states in real nanophononic systems, we performed all-optical Raman scattering measurements on a planar GaAs/AlAs sample with a layer structure as sketched in Figure 3a. The sample was grown by molecular beam epitaxy (MBE) on a (001) GaAs substrate and consists of two parts: In the center it contains two concatenated acoustic DBRs with 20 GaAs/AlAs bilayers each, designed for a frequency of $2f_0$ = 354 GHz and with inverted bands corresponding to the parameter choice $\delta = -0.1$ and $\delta = +0.1$ for the left (right) DBR, respectively. This acoustic structure is enclosed by two GaAlAs-based optical DBRs (see Methods section for details) designed for a wavelength of $\lambda_{opt}$ = 940 nm such that the acoustic nanostructure serves as a $2\lambda_{opt}$ spacer of a resonant optical microcavity [50–52].

The resonance energy of a planar optical microcavity shows a parabolic dependence on the incidence angle. It is therefore possible to establish the Double Optical Resonance (DOR) with a single cavity mode by tuning the laser wavelength to match the resonance wavelength at a given incidence angle, and resonantly collecting the scattered Raman signal at normal incidence (see the schematics in Fig. 3a) [19,50]. Moreover, the use of the optical microcavity allows us to access phonons usually observable under back and forward scattering geometries, while at the same time enhancing the signals by several orders of magnitude [52].

In practice, instead of tuning the laser to match the cavity mode, the sample was grown with an in-plane thickness gradient, such that its optical resonance shifts with a gradient of approximately 100 nm per inch across the wafer. This allows us to fix the laser wavelength and establish the DOR condition by only optimizing the position on the sample. The angle of incidence, in addition, allows us to select the frequency range of the Raman signal. Note that the resonance frequency of the topological interface state also shifts due to the in-plane thickness gradient of the sample. An experimental scan in position with a fixed angle of incidence is shown in Figure 3b. The shift of the interface state amounts to approximately 600 MHz over the displayed region. As a function of the incidence position we observe that the collected Raman intensity varies substantially over few-micrometer distances and exhibits a clear maximum which corresponds to simultaneous resonant excitation and collection for Raman photons at frequency shifts around 350 GHz.

A single Raman spectrum measured under DOR condition is displayed in Figure 3c (black). Three clear peaks at 323GHz (A), 360GHz (B) and 397GHz (C) can be observed. As discussed below, peak B corresponds to the topological interface mode, while peaks A and C are phonons distributed along the DBRs generating Raman signals in backscattering configuration. These peaks are a general feature of acoustic superlattices and samples formed by DBRs [53–56].

The simulated acoustic reflectivity of the studied sample is shown in Figure 3d. The clear dip in the stopband around 360GHz is generated by the topological interface mode between the two DBRs and can be assigned to the main peak (B) of the experimental spectrum. Notice that if the spatial mode symmetries of any of the two constituent DBRs are changed, this interface mode disappears and the reflectivity would simply show the stop band.

In Figure 3e the layer structure of the acoustic part of the sample is superimposed with the calculated mechanical displacement patterns at the three peak frequencies. Profile B presents a characteristic exponential decay into the DBRs. This decay is determined by the reflectivity of the two DBRs. Peaks A and C, on the contrary, show an almost uniform displacement along the structure, without any indication of confinement. These modes extending over the full structure fulfill the resonance condition for Raman active backscattering in periodic structures [46]. To further account for the experimental results, we performed a photoelastic model calculation[19,46,53] based on Eq. (5).

$$I(\omega) \propto \left| \int_{-\infty}^{\infty} E_{laser}(z) E_{scat}^*(z,\omega) \frac{\partial u(z,\omega)}{\partial z} p(z) \mathrm{d}z \right|^2 \qquad (5)$$

In this model the frequency-dependent Raman intensity $I(\omega)$ is simulated by evaluating the mode overlap between the electric field of the incident laser, the outgoing Raman scattered field and the strain distribution of the corresponding phonon at frequency $\omega$. The integral furthermore contains the material- and wavelength-dependent photoelastic constant $p(z)$. In our case with an experimental wavelength of $\lambda_{opt} = 915$ nm only the photoelastic contribution of GaAs with an electronic transition around 850 nm needs to be taken into account. The photoelastic constant of AlAs is negligible since the closest electronic transition occurs at much higher energies. The results of the simulation are plotted together with the experimental data in Figure 3c (solid red). We have furthermore taken into account the finite resolution of the spectrometer of 13 GHz by convoluting the simulation with a Gaussian distribution (dashed red). Evidently, the model captures all features of the measured data. In particular, the relative heights and spectral positions of the three peaks are in excellent agreement. We stress that for this model we only used design parameters of the sample. No further fitting was needed apart from global scaling factors to account for the in-plane thickness gradient of the sample and the measured overall intensity.

In what follows, we describe how the implementations described in Figures 2 and 3a are related. The system in Figure 2 is based on DBRs composed of centro-symmetric unit cells, a conception coming from topology. The systems in Figure 3 with DBRs composed of integer numbers of bilayers is the paradigm for standard solid state microoptical and nanophononic devices. Figure 4 shows a series of calculated reflectivity spectra corresponding to concatenated pairs of DBRs with inverted bands where the DBRs are formed by an integer number of centro-symmetric unit cells (red) or an integer number of GaAs/AlAs bilayers (blue). In what follows we discuss how to map these two cases.

We start from standard bilayers as sketched in the top part of **Figure 4**a. The interface is formed by a light green (AlAs, $\delta = -0.1$) layer and dark blue (GaAs, $\delta = +0.1$) layer. This is the same system as studied in Figure 3. Half a dark blue layer equals half a dark green layer plus an additional remainder. This thin remainder has a thickness of $\Delta d_{GaAs} = \frac{v_{GaAs}}{4f_0}\delta$, which is much smaller than each of the other layers in the full structure. By performing this decomposition which is graphically illustrated in case II) of Figure 4a, the interface between the two DBRs can now be set between two DBRs constituted by centro-symmetric unit cells separated by a perturbation (orange) that can be adiabatically varied in thickness. When this perturbation is absent, the system reproduces the case discussed in Figure 2. This case is sketched in case III) of Figure 4a.

By following this method it is always possible to associate a system based on an integer number of standard bilayers to a system based on centro-symmetric unit cells. There is a smooth evolution of the mode frequency between the two extreme cases (Fig. 4b, black curves), in other words it is always possible to go smoothly from configuration I) to configuration III) in Figure 4a. The existence of a confined state in configuration I) implies the existence of an associated mode in the adiabatically connected situation III) of Figure 4a. Two direct consequences arise: 1) The mode in configuration I) will be slightly red-shifted with respect to the associated mode generated by centro-symmetric unit cells in case III) since an effective small propagation phase $\phi_{spacer}$ is introduced for the bilayers. In the studied case the mode appears shifted by only approximately 2 GHz for a frequency of 350 GHz. This shift is hence well below our experimental resolution. 2) The existence of a mode in the centro-symmetric configuration is not a sufficient condition for the existence of an associated mode in configuration I). The extreme case of a topologically confined mode vanishing when changing to the configuration of standard bilayers however only occurs for values of $\delta$ in a small range around $\delta \approx \pm 0.5$. The emergence of this second consequence is addressed in detail in Appendix A and B.

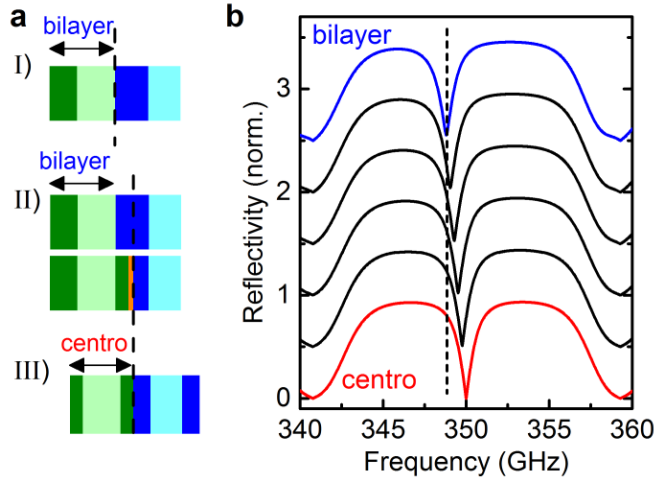

**Figure 4 Symmetrization of the topological interface state; a**, (top) Interface between two DBRs constructed from standard bilayers used in the experimental configuration. (middle) We remove a thin layer of GaAs at the interface (black), such that the rightmost layer of the left DBR becomes as thick as the leftmost layer of the right DBR (i.e. we replace a dark blue by a dark green layer). (bottom) Resulting interface between two topologically different DBRs constructed from centro-symmetric unit cells. **b**, Corresponding phonon reflectivity spectra of two concatenated DBRs from different topological phases with 20 unit cells each ($\delta = -0.1$ on the left and $\delta = +0.1$ on the right). From top to bottom we gradually tune the unit cells from standard asymmetric bilayers (blue) to centro-symmetric unit cells (red) by removing GaAs at the interface resulting in a small perturbation of the mode frequency. For better visibility subsequent curves are vertically offset by 0.5.

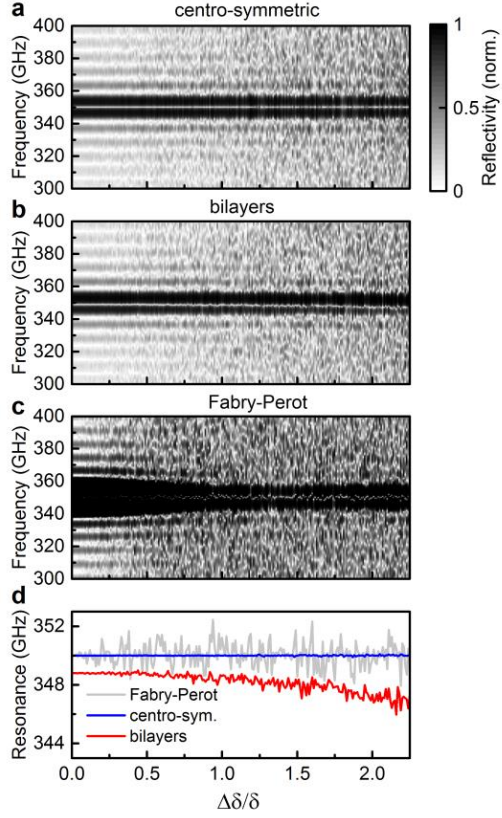

**Figure 5: Stability of the interface state to disorder.** Simulated phonon reflectivity of two concatenated DBRs with inverted bands (δ=-0.1 first DBR, δ=+0.1 second DBR) each with 20 unit cells. A dip in reflectivity indicates the presence of a topological interface mode at 350GHz (cf. also Figure 2b). Random fluctuations in δ for each unit cell are introduced with a uniform distribution ranging from $-\Delta\delta/\delta$ to $+\Delta\delta/\delta$. **a,** Centro-symmetric unit cells **b,** Standard bilayers **c,** Fabry-Perot resonator (see text for details). **d,** Resonance frequency of the confined mode as a function of disorder strength Δδ/δ. The Fabry-Perot resonance undergoes fluctuations that are much stronger than the ones shown by the topological cavities. The acoustic minigap in panels **a** and **b** ranges from 347GHz to 353GHz while it ranges from 340GHz to 360GHz in panel **c**, respectively.

An important property of topologically protected states is their robustness against disorder that does not change the underlying topological invariants. In what follows, we evaluate the robustness of the investigated structures. A disorder that does affect the Zak phases can be introduced as a uniform distribution of variations in δ (ranging from -Δδ/δ to +Δδ/δ) for each unit cell. We compare the performance of three devices: i) Two concatenated DBRs, each one composed of an integer number of centro-symmetric unit cells with inverted bands (δ=-0.1 first DBR, δ=+0.1 second DBR), i.e. the structure presented in Figure 2. ii) Two concatenated DBRs, each one composed of an integer number of standard bilayer unit cells with inverted bands (δ=-0.1 first DBR, δ=+0.1 second DBR), i.e. the structure that was reported in Figure 3. iii) A standard Fabry-Perot resonator formed by two identical DBRs ($\lambda/4$, $3\lambda/4$ corresponding to $\delta = 0.5$ which maximizes the span of the stop band) enclosing a $\lambda/2$ spacer.

In Figure 5a-c we show simulated phonon reflectivity spectra for the three structures as a function of the disorder strength Δδ/δ. In panel d we plot the dependence of the resonance frequency for each of the confined phonon modes. In panels a-c the black areas correspond to the stop band, the oscillations on the side of the stop band correspond to Bragg oscillations. As **Δδ/δ** increases, these Bragg oscillations disappear. The line within the stop band corresponds to the confined phonon mode.

As shown in panel d, for the centro-symmetric case (blue) the mode remains stable for all values of Δδ/δ and well centered in the minigap. The topological interface state between two DBRs formed by bilayers (red) remains stable around the unperturbed frequency for $\frac{\Delta\delta}{\delta} < 1$. For bigger values it presents a clear red-shift. In contrast, the frequency of the Fabry-Perot cavity mode (grey) remains centered in the minigap, but it undergoes fluctuations that well exceed those observed for the bilayer case. We observe that the topological structures outperform the standard Fabry-Perot resonator in stability. It must be noted, however, that this is not an indication for the quality factor of the structures.

For the first structure, the perturbations on the unit cell do not affect the Zak phases corresponding to the two DBRs provided that $\frac{\Delta\delta}{\delta} < 1$. As a consequence, the mode remains pinned at the center of the minigap. In the second case, a change in δ affects the Zak phase and thus results in frequency fluctuations and an overall shift of the interface mode. For the third case, the stability of the mode is mainly determined by fluctuations of the thickness in the central spacer. Notice that for this case, regardless of the magnitude of the perturbation, there is always a confined mode within the stop band. The topological interface states between two different DBRs are more tolerant to the explored thickness fluctuations than the standard Fabry-Perot resonator formed by two identical DBRs separated by a spacer.

**Conclusions**
In summary, by applying the concept of band inversion to nanophononic periodic superlattices we have successfully constructed a topological nanophononic interface state at 350GHz. Contrary to a Fabry-Perot resonator, where two identical DBRs enclose a resonant spacer, the implemented resonator relies on two different DBRs without any spacer. We experimentally evidence the existence of the topologically confined mode by high resolution Raman scattering spectroscopy. Corresponding photoelastic model calculations perfectly account for all major features of the measured Raman spectra, in particular the signature peak of the topological interface state at approximately 350GHz.

We have calculated Zak phases for the bands bounding the considered minigap, which are directly associated to the reflection phases of the individual superlattices. In the case of a DBR terminated by a centro-symmetric unit cell, the Zak phase purely depends on the sign of δ, and becomes a good topological number to engineer a confined state. Structures based on DBRs terminated by bilayers can be considered as small perturbations of the centro-symmetric case. As such, the same Zak phases used for the structure terminated by a centro-symmetric unit cell can be used for a wide range of values of δ to generate a topologically confined mode. It is worth mentioning that for the perturbed case the Zak phases computed using Eq. (3) do not show a discrete distribution of just two possible values 0 and $\pi$.

The presented GaAs/AlAs material platform is at the base of a wide range of applications in opto-electronics, photonics and nanophononics. It is also naturally compatible with active media. The discussed construction principle and material platform can be directly applied in the 20GHz range where full control of the 3D phononic density of states and strong opto-mechanical interactions were recently demonstrated [10,11]. Since we have established a direct connection between centro-symmetric unit cells and standard bilayers these concepts can be readily transposed to existing real-life applications in opto-electronics, photonics and opto-mechanics.

This work bridges two research fields; topology and nanomechanics. On one side, we show how acoustic phonons can constitute a platform to study topological properties. On the other side, the use of topological invariants makes it possible to revisit the problem of phonon confinement with exciting perspectives.

**Methods**

Sample preparation: The sample was grown by molecular beam epitaxy (MBE) on a (001) GaAs substrate. The outer optical cavity DBRs were grown from alternating layers of $Ga_{0.9}Al_{0.1}As/Ga_{0.05}Al_{0.95}As$ with an optical thickness of $\lambda_{opt}/4$ per layer at a vacuum wavelength of $\lambda_0 = 940$nm and 12(16) layer pairs on the air (substrate) side. The topological acoustic structure between the optical DBRs was grown from alternating layers of GaAs/AlAs with 20 layer pairs for the DBR facing air and 19.5 layer pairs facing the substrate. Layer thicknesses are given by $d_{GaAs} = \frac{v_{GaAs}}{4f_0}(1+\delta)$ and $d_{AlAs} = \frac{v_{AlAs}}{4f_0}(1-\delta)$ with $\delta_{left} = -0.1$ (DBR facing air) and $\delta_{right} = 0.1$ (DBR facing substrate) for a design frequency of $2f_0 = 354.2$GHz using the values $v_{GaAs} = 4780$m/s and $v_{AlAs} = 5660$m/s for room temperature. The parameters are chosen such that the optical path length of the acoustic structure is exactly $2\lambda$, i.e. in the optical domain the acoustic structure represents the resonant spacer of a cavity. Compared to the bare acoustic structure this configuration has two main advantages: First, the resonant enhancement of both the incident excitation laser field and the scattered Raman field leads to an increase in observable Raman intensity by up to five orders of magnitude in double optical resonance (DOR) configuration. Second, the selection rules for forward and backward scattering Raman signals are lifted such that both types of signals become accessible in the backward scattering configuration that we implemented here. The sample was furthermore grown with an in-plane thickness gradient such that its optical resonance varies from 830nm-1050nm under normal incidence across a 2″ wafer. This gradient allows us to keep the optical wavelength of the excitation laser fixed and establish the DOR condition by optimizing both the position and incidence angle on the sample.

Raman measurements: Raman scattering experiments were performed at room temperature in backscattering configuration. For optical excitation we used a tunable continuous wave Ti:Sapphire laser (M2 SolsTiS) working at a wavelength of 915nm. We irradiated an approximate power of 20mW onto the sample surface at an incidence angle of 11° and focalized to a 50μm spot. Raman spectra were collected normal to the sample surface and recorded with a liquid nitrogen cooled CCD camera (Princeton Instruments) after being dispersed in a double monochromator (HIIRD2 Jobin Yvon). To establish the DOR condition we optimized both the incidence angle of the laser on the sample and the position on the sample along the in-plane thickness gradient.

**Appendix A: Existence of an interface state in the acoustic minigap**

When considering concatenated DBRs composed of standard GaAs/AlAs bilayers, a counter-intuitive dependence of the topological state on the width of the acoustic minigap occurs. In a standard acoustic Fabry-Perot resonator, where two superlattices enclose a spacer, the broader the bandgap, the shorter the evanescent confinement length of the cavity mode in the DBRs [10,57]. That is, the spatial confinement can be optimized by maximizing the width of the acoustic minigap. However, in the case of a topological nanophononic mode confined in between two DBRs made from standard bilayers, an increase in the width of the minigap can induce its complete disappearance. We illustrate this effect in Figure 6 by considering again the example of two DBRs with 20 unit cells each and inverted bands bounding the minigap centered at $2f_0 = 350$ GHz. In Figure 6 we vary the parameter $\delta$, which determines the width of the minigap, and study its influence on the resonance of the topological interface state. In panel a we show the upper and lower band edge (blue and red) together with the resonance frequency of the topological nanophononic interface state for unit cells from standard bilayers (black) and for centro-symmetric unit cells (grey, c.f. Figure 2 of the main text). For this calculation we always consider two DBRs for which the condition $-\delta_{left} = \delta_{right}$ is satisfied. As an example, we plotted the phonon reflectivity for $\delta = -0.2$ (left DBR) and $\delta = +0.2$ (right DBR) in panel b, as indicated by the dotted vertical line in panel a. Like in Figure 4 of the main text, we observe that the frequency of the topological state is redshifted from the bandgap center in the case of standard bilayers. With increasing values of $\delta_{right}$ we observe that the redshift grows nonlinearly. In particular, for a value of $-\delta_{left} = \delta_{right} \approx 0.4$ the resonance occurs exactly on the band edge. For values beyond this point no resonant interface state is found inside the bandgap anymore. In contrast, for the centro-symmetric unit cells the topological interface state always appears at the center of the acoustic minigap (grey). Hence, the wider we open the acoustic bandgap, the more susceptible the confined mode becomes to the small perturbation of the interface that we introduce by changing from centro-symmetric unit cells to bilayers.

As we show in Appendix B, all the observations stated above are fully backed up by taking into account the phase shifts due the addition of a small perturbation at the interface between two superlattices as depicted in Figure 4 (see main text).

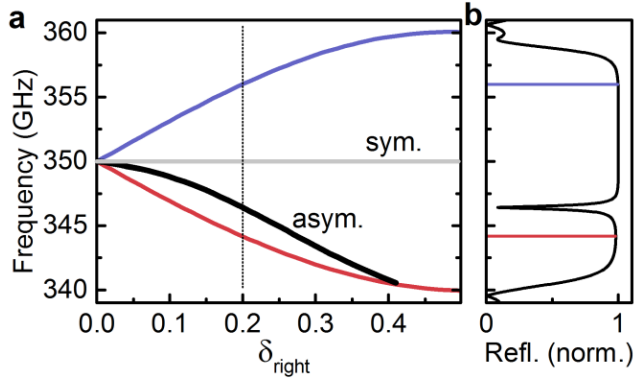

**Figure 6** Topological interface state resonance tuned by width of the bandgap. **a**, Upper and lower band edge (blue and red) as a function of parameter $\delta$. We show the resonance frequency of a topological interface state (black) confined in between two DBRs with $\delta_{left} = -\delta_{right}$, each made from 20 standard GaAs/AlAs bilayers. With growing band gap the interface state redshifts towards the lower band edge and ceases to exist beyond $\delta \approx 0.4$. **b**, Phonon reflectivity spectrum for $\delta_{right} = 0.2$ (indicated by the dotted line in panel a). The horizontal blue and red line indicate the width of the phononic band gap for this choice of parameters, the sharp dip in reflectivity indicates the resonance frequency of the interface state.

## Appendix B: Finite size effects on the topological state frequency

A detailed understanding of the precise evolution of the interface state resonance inside the minigap requires to take into account the actual evolution of the reflection phases $\phi_{left}$ and $\phi_{right}$ across the acoustic band gap. We show that these phases are a function of the system size, i.e. the number of DBR unit cells that we consider.

Considering centro-symmetric unit cells, an interface mode appears at the center of the acoustic minigap when concatenating two DBRs with inverted bands [37,38]. Figure 7a illustrates the two particular choices of centro-symmetric unit cells (left) and standard GaAs/AlAs bilayers (right) as discussed in the experimental section of the main text. The color codes remain the same as in Figure 2 of the main text. As a first step to formalize the perturbation of the interface between two DBRs by an additional layer, we add a new parameter $0 \leq \kappa \leq 1$ to the parametrization of the unit cell by $\delta$, as introduced in Figure 1a. $\kappa$ quantifies the internal distribution of the materials inside the unit cell when changing gradually from centro-symmetric unit cells to standard bilayers. More precisely, for $\kappa = 0$ ($\kappa = 1$) the unit cell starts with a full layer of AlAs (GaAs) followed by GaAs (AlAs) and for $\kappa = 0.5$ a full layer of AlAs is enclosed in between two GaAs half-layers, i.e. the unit cell is centro-symmetric. While $\kappa$ does not have any influence on the band structure of an infinite periodic superlattice, it evidently has a critical impact on the terminating layer of a DBR, i.e. on the reflection phases $\phi_{left}$ and $\phi_{right}$.

When inspecting the concatenated DBRs on the left side of Figure 7a on the level of material layers instead of unit cells, it becomes evident that the structure may also be regarded as a standard $\lambda/2$ cavity spacer resonant at $2f_0$, enclosed in between two DBRs with a phononic bandgap centered at $2f_0$. To see this, take into account that the two layers bounding the interface in between the two DBRs contribute acoustic path lengths of $0.5\frac{v_{GaAs}}{4f_0}(1+\delta)+0.5\frac{v_{GaAs}}{4f_0}(1-\delta)=0.5\frac{v_{GaAs}}{2f_0}$ resulting in a combined half-wavelength central layer of GaAs.

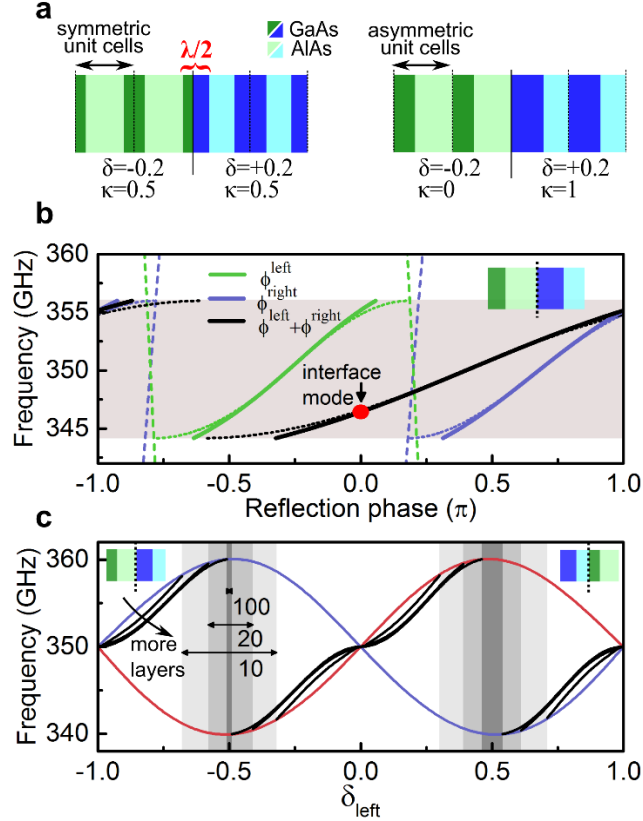

**Figure 7 Finite size effects on the topological state frequency, a**, Interface between two DBRs with inverted bands for centro-symmetric unit cells (left) and standard bilayers (right). All color-codes as in the main text. **b**, Reflection phases $\phi_{left}$ (green) and $\phi_{right}$ (blue) as well as their sum (black) across the acoustic minigap (shaded grey) for $\delta_{left}$ = -$\delta_{right}$ = -0.2 and bilayer unit cells (solid lines: 20 unit cells per DBR, dotted lines: semi-infinite DBRs). The red dot marks the frequency of the topological interface state confined between these two DBRs. Dashed lines show phase shifts following Eq. (6). **c**, Topological interface state resonance frequency as a function of parameter $\delta$ for two DBRs with bilayer unit cells and $\delta_{left}$ = -$\delta_{right}$. The band edges are shown in red and violet (indicating the edge mode symmetries of the left DBR). The interface state resonance frequencies for 10, 20 and 100 bilayers in each of the two concatenated DBRs are shown in black. Parameter ranges for which no interface state occurs are shaded in grey, numbers indicate the number of unit cells per DBR.

By comparing this cavity-like configuration to the bilayer configuration on the right of Figure 7a, it is however possible to establish a direct mapping between them. Consequently, we can use the symmetric configuration as a starting point to investigate the resonance conditions for the interface configuration which we have implemented experimentally. Essentially, this mapping consists in removing half a layer of GaAs from the rightmost unit cell of the left DBR and adding half a layer of GaAs to the leftmost unit cell of the right DBR, following the lines of Figure 4a in the main text. These two layers are however of different thickness due to the band inversion, i.e. due to the different value of $\delta$ on the left and on the right lattice. In Figure 4 the particular case of going from centro-symmetric unit cells to bilayers was illustrated, furthermore satisfying the condition $\delta_{\text{left}} = -\delta_{\text{right}}$. In the general case of an arbitrary choice of κ and δ the additional reflection phase of a semi-infinite DBR that has to be taken into account compared to the reflection phase of a DBR composed of centro-symmetric unit cells is given by the family of lines

$$\Phi_m(f, \kappa, \delta) = -\pi\left[(0.5 - \kappa)(1 + \delta)\frac{f}{f_0} - m\right], \quad m \in \mathbb{Z} \tag{6}$$

In Figure 7b we show the reflection phases $\phi_{left}$ (green) and $\phi_{right}$ (blue) for the nanophononic interface state. Here, we chose $\delta = -0.2$ on the left and $\delta = +0.2$ on the right and DBRs composed of 20 standard bilayers. For frequencies inside the acoustic minigap (shaded in grey) we have calculated the individual reflection phases (green and blue) and their sum (black), as well as the reflection phases for a pair of semi-infinite DBRs with the same parameters for κ and δ (dotted, same colors). The dashed green and blue lines indicate the corresponding frequency-dependent phase shifts $\Phi_m$ following Eq. (6). We find that the resonance condition for the interface state is fulfilled at a frequency of 346 GHz (marked by the red dot) and that taking the additional phase shifts into account results in the interface mode resonance being offset from the bandgap center. To arrive at a more complete picture of the conditions under which the band inversion principle allows the confinement of an acoustic interface state in the common band gap of two concatenated DBRs, we furthermore systematically varied δ and traced the evolution of the interface state resonance for different numbers of unit cells. The results are summarized in Figure 7c. As a function of $\delta_{left} = -\delta_{right}$ we show the two edges of the second acoustic band gap (red and blue) and determined the resonance frequency of the resulting acoustic interface state when concatenating two DBRs with 10, 20 and 100 unit cells each (black).

We find that for all numbers of unit cells interface states appear for very large and very small magnitudes of δ, but as we approach $\delta \approx \pm 0.5$, i.e. a maximally opened acoustic band gap, the resonances move closer to the band edges and eventually cease to exist beyond a critical magnitude (areas shaded in grey). We also observe that these critical points lie closer to the maximally opened gap for a larger number of unit cells. That is, a larger pair of DBRs supports acoustic interface states for a wider range of δ. The origin of this trend becomes clear from the observation in Figure 7b, that for a finite DBR the reflection phase does not span a full interval of π over the course of the band gap. Hence, for a larger pair of DBRs the resonance condition from Eq. (2) can still be fulfilled closer to the band edges.

**Corresponding Author**

* Correspondence should be addressed to N.D.L.K. daniel.kimura@c2n.upsaclay.fr

**Author Contributions**
M.E. and F.R.L. contributed equally to this work.

All authors substantially contributed to this work. F.R.L. and M.E. conducted the experiments. M.E. performed the theoretical calculations. A.L. and C.G.C. fabricated the samples. N.D.L.K. and A.L. proposed the concept. N.D.L.K. guided the research.

**Acknowledgements**

The authors acknowledge funding through the ERC Starting Grant No. 715939 Nanophennec, through a public grant overseen by the French National Research Agency (ANR) as part of the "Investissements d'Avenir" program (Labex NanoSaclay, reference: ANR-10-LABX- 0035), funding by the ANR (reference: QDOM: ANR-12- BS10-0010) and by the French RENATECH network.

**Competing financial interest:**

The authors declare no competing financial interests.